\documentclass[aps,prl,numerical,linenumbers,superscriptaddress,showpacs,floatfix,reprint]{revtex4-1}

\usepackage[dvipdfm]{graphicx,color}
\usepackage{amsmath,amssymb,amsfonts}
\usepackage[colorlinks=true,linkcolor=blue,plainpages=false,hypertex]{hyperref}

\newcommand{\be}{\begin{equation}}
\newcommand{\ee}{\end{equation}}
\newcommand{\ba}{\begin{eqnarray}}
\newcommand{\ea}{\end{eqnarray}}
\newcommand{\ban}{\begin{eqnarray*}}
\newcommand{\ean}{\end{eqnarray*}}
\newcommand{\nn}{\nonumber}

\def\v2{\mbox{$v_2$}}

\begin{document}

\title{A new method for the experimental study of topological effects \\ 
in the quark-gluon plasma}
\medskip

\author{N. N. Ajitanand}
\affiliation{Department of Chemistry, 
Stony Brook University, \\
Stony Brook, NY, 11794-3400, USA}
\author{Roy A. Lacey}
\affiliation{Department of Chemistry, 
Stony Brook University, \\
Stony Brook, NY, 11794-3400, USA}
\author{A. Taranenko}
\affiliation{Department of Chemistry, 
Stony Brook University, \\
Stony Brook, NY, 11794-3400, USA}
\author{J. M. Alexander}
\affiliation{Department of Chemistry, 
Stony Brook University, \\
Stony Brook, NY, 11794-3400, USA}
\date{\today}

\begin{abstract}
A new method is presented for the quantitative 
measurement of charge separation about the reaction plane.
A correlation function is obtained whose  
shape is concave when there 
is a net separation of positive and negative charges.
Correlations not specifically associated with charge, from 
flow, jets and momentum conservation, do not influence the shape or 
magnitude of the correlation function.
Detailed simulations are used to demonstrate the effectiveness of the method 
for the quantitative measurement of charge separation. Such measurements   
are a pre-requisite to 
the investigation of topological charge effects in the QGP as
derived from the ``strong $\cal{CP}$ problem''.
\end{abstract}

\pacs{Valid PACS appear here}
\maketitle

\section{Introduction}
Topological charge fluctuations play an important role in the structure 
of the QCD vacuum \cite{Schafer:1996wv}. It manifests in the breaking of 
chiral symmetry, as well as in the mass spectrum and other properties of hadrons.
These fluctuations can also lead to the formation of metastable vacuum 
domains, especially in the vicinity of the de-confinement 
phase transition, in which fundamental symmetries (${\cal P}$ and/or $\cal{CP}$) are 
spontaneously broken \cite{Kharzeev:2009fn} {\em i.e.} the so-called 
``strong $\cal{CP}$ problem''. Experimental evidence for such topological fluctuations 
have been largely indirect.

Recently, it has been suggested that direct experimental signatures of  
topological fluctuations could result from quark gluon plasma (QGP)  
[quarks liberated from hadronic confinement]
subjected to an intense (hadron-scale) external magnetic field, 
via the so called "chiral magnetic effect" (CME) \cite{Kharzeev:2004ey,Fukushima:2009ft}. 
In brief, topological charge fluctuations in the QGP leads to an axial anomaly or local 
imbalance between left-handed and right-handed light quarks. 
In an intense magnetic field, these quarks move along the field
to create a net electric current which results in a separation of positive 
and negative electric charges in the field direction. 
Evidence for the chiral magnetic effect has been found in recent 
numerical lattice QCD calculations \cite{Buividovich:2009my}.
An axial anomaly can also result from an anomalous global symmetry 
current in the hydrodynamic description of the QGP \cite{Son:2009tf}. This 
results in a modification of the hydrodynamic current by a term proportional 
to the vorticity of the fluid, and manifests also as a separation of positive 
and negative electric charges perpendicular to the reaction plane. Hereafter,
we term this as the chiral rotation effect (CRE).

\section{Measuring topological effects}

\begin{center}
\begin{figure}[t]
\includegraphics[width=1.0\linewidth]{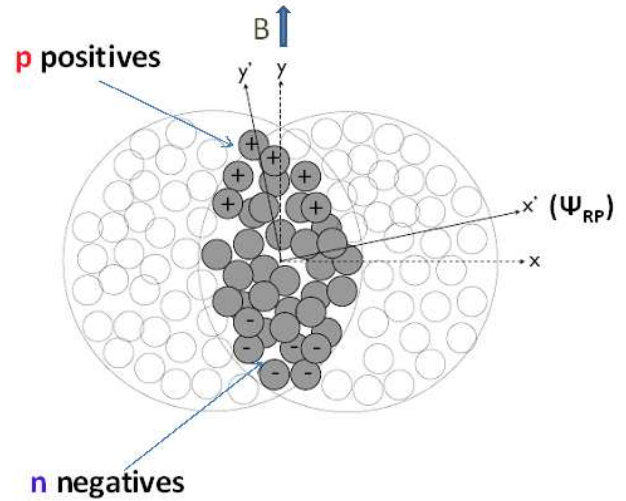}
\caption{Schematic illustration of the chiral magnetic effect. Colliding nuclei (depicted as circles) are
moving in and out of the page respectively. The Magnetic field ($B$) and system orbital angular momentum ($L$) are
perpendicular to the reaction plane (plane spanned by the impact parameter, b, and colliding nuclei
direction). Note that the overlap zone is not necessarily aligned with $B$ or $L$.
}
\label{Fig1}
\end{figure}
\end{center}

	Collisions between heavy nuclei at the Relativistic Heavy Ion Collider 
(RHIC), not only create a strongly coupled low viscosity 
QGP \cite{Adcox:2004mh,Lacey:2006bc,Luzum:2008cw,Song:2008hj,
Chaudhuri:2009hj,Lacey:2009xx,Denicol:2010tr,Lacey:2010fe} but also the strongest 
magnetic fields [orthogonal to the reaction plane] attainable 
in the laboratory \cite{Rafelski:1975rf}. 
Consequently, the chiral magnetic effect or the chiral rotation effect 
is expected to lead to a charge asymmetry in the distribution of particles 
emitted about the reaction plane 
(see Fig. \ref{Fig1}). Experimental studies of such an asymmetry could provide 
an important avenue for investigating one of the most important problems of strong 
interaction theory.

	The STAR collaboration has analyzed data from recent measurements of Au+Au 
and Cu+Cu collisions (at $\sqrt{s_{NN}} = 200$ GeV) in search of this 
charge asymmetry with respect to the reaction plane. 
To do this they constructed a correlator which used the emission 
angles of like-sign ($++ \text{or} --$) and 
opposite-sign ($+\,-$) hadron pairs. The correlator is defined by the 
event average
\be 
\label{eq-CSTAR}
C^{(\pm,\pm)} = \left\langle \cos\big(\phi_\alpha^{(\pm)} + 
\phi_\beta^{(\pm)} -2 \Psi_{\rm 2}\big) \right\rangle,
\ee
where $\phi_\alpha,\phi_\beta$ denote the azimuthal emission angles of any pair 
of hadrons, and $\Psi_{\rm 2}$ denotes the azimuthal 
orientation of the estimated second order event plane. 
The difference
\be
\Delta Q \sim C^{(++)}+C^{(--)} - 2C^{(+-)}
\ee 
was used to test for a charge separation [about the reaction plane] of the kind suggested 
by the CME and the CRE, after an appropriate correction for dispersion of the 
reaction plane. 

	A charge separation has been reported by the STAR 
collaboration \cite{Abelev:2009uh,Abelev:2009txa}. However, its mechanistic origin is
still under intense debate \cite{Asakawa:2010bu,Bzdak:2010fd,Muller:2010jd,
Wang:2009kd,Schlichting:2010na}. One reason for this has been the observation that the 
correlator used in the STAR analysis may be sensitive to several well known 
``background'' correlations such as elliptic flow, jets and momentum 
conservation \cite{Bzdak:2010fd,Wang:2009kd,Schlichting:2010na}. 
Therefore, it is important to develop and investigate new 
correlators which can overcome many, if not all, of these deficiencies.

	A full study of topological effects in the QGP and its implications for the 
``strong $\cal{CP}$ problem'', will undoubtedly require further detailed measurements 
focused on accurate experimental quantification of the dependence of charge asymmetry on 
particle species, particle $p_T$, collision-system deformation, event centrality and beam 
collision energy. Here, we present a new experimental correlator specifically designed to 
aid such investigations.

	Our technique involves a multi-particle charge-sensitive  
in-event correlator $C_c (\Delta S)$,which is expressed as a ratio of 
two distributions;
\be
C_c (\Delta S) = \frac{N(\Delta S_{csep})}{N(\Delta S_{cmix})}.
\label{eq:1}
\ee 
The numerator is a distribution over events of the event averaged 
quantity $\Delta S_{csep}$ defined as  
\be
\Delta S_{csep} = \left\langle {S_p^{h + }} \right\rangle  - \left\langle {S_n^{h - }} \right\rangle  
\label{eq:2}
\ee
where  
\ba
\left\langle {S_p^{h + }} \right\rangle  = \frac{{\sum\limits_1^p {\sin (\Delta {\varphi _ + })} }}{p}, \,\,\,
\left\langle {S_n^{h - }} \right\rangle  = \frac{{\sum\limits_1^n {\sin (\Delta {\varphi _ - })} }}{n}, 
\label{dels}
\ea 
$n$ and $p$ are the numbers of negative and positive hadrons 
[respectively]
emitted about the observed event plane $\Psi_{EP}$ ($m = n +p$ is 
the 
charge hadron multiplicity for an event)  
and $\Delta {\varphi}= \phi - \Psi_{EP}$ where $\phi$ is the azimuthal 
emission angle of the charged hadron.

%
%

The distribution $\Delta S_{cmix}$ in the denominator in Eq. 3, is 
obtained by making event averages in a slightly different way;
That is, Eq. 5 is used to evaluate the averages 
$\left\langle {S_p^{h}} \right\rangle$ and $\left\langle {S_n^{h}} 
\right\rangle$ 
for $p$ and $n$ randomly chosen hadrons (irrespective of charge) i.e.
\be
\Delta S_{cmix} = \left\langle {S_p^{h}} \right\rangle  - \left\langle {S_n^{h}} \right\rangle. 
\label{eq:6}
\ee

	There are several important features of the new correlator $C_c 
(\Delta S)$. First, 
it is constructed entirely from a real event; hence, it is pure in event 
class (centrality, vertex, etc). Second, it is rather insensitive to 
the background correlations which influence reliable extraction of the magnitude 
of the charge-separation correlation (see discussion below). In what follows, we 
use detailed simulations to demonstrate the expected trends, as well as the efficacy 
of $C_c (\Delta S)$.

\section{Simulation Methodology}

The response of $C_c (\Delta S)$ to a charge-separation signal was  
tested via a detailed set of simulations tuned to reproduce 
observed experimental features. 
The simulations included the following major steps for each event. 
\begin{itemize}
\item The event plane was chosen at random from $2\pi$.
Charged particles were then emitted with an azimuthal distribution with 
respect to this 
reaction plane as:
\ba
N(\Delta {\varphi}) \propto (1+2v_2\cos\Delta {\varphi})+2v_4\cos(4\Delta {\varphi}) \nn \\
 + 2a_1\sin(\Delta {\varphi}),
\ea
where the Fourier coefficients $v_2$ and $v_4$ are the observed  
magnitudes of elliptic and hexadecapole flow, and $a_1$ is 
the charge-separation signal of interest. The number and  $p_T$ 
distribution of particles were tuned to match the experimentally 
observed distributions. The reaction plane was then dispersed according 
to the experimentally observed dispersion for the centrality selection 
under study. 
\item Neutral decay particles (e.g. $\Lambda$ and $K_0$) were emitted 
with respect to the reaction plane according to their observed flow 
patterns.
The decay kinematics of these resonances were followed so as to obtain 
the daughter particle directions and momenta. The relative abundance of the 
decay particles were constrained by the requirement that the simulated and observed 
positive-negative charge pair correlations (obtained by the 
standard event mixing method) were in agreement.
\item Jet particles were emitted with respect to the jet axes in a 
manner which was consistent with the observed two-particle jet 
correlations.
\item All emitted particles were passed through an acceptance filter specifically
designed to take account of the detector acceptance and consequently, reproduce 
the measured inclusive distributions for positive and negative hadrons 
respectively as a function of $p_T$.
\item The simulated events were analyzed as if they were actual experimental 
events. The correlation $C_c (\Delta S)$ was evaluated for the selected 
range of $p_T$ using the detected particles and the 
dispersed reaction planes using Eqs. \ref{eq:1} - \ref{eq:6}.

\end{itemize}

\section{Results from simulations}

Simulations were performed for a broad spectrum of scenarios. Here, we 
show a representative set of results which lends insight into the detailed 
nature of $C_c (\Delta S)$, as well as its sensitivity to different  
sources 
of background correlations. 

\begin{figure}
\includegraphics[scale = 0.45]{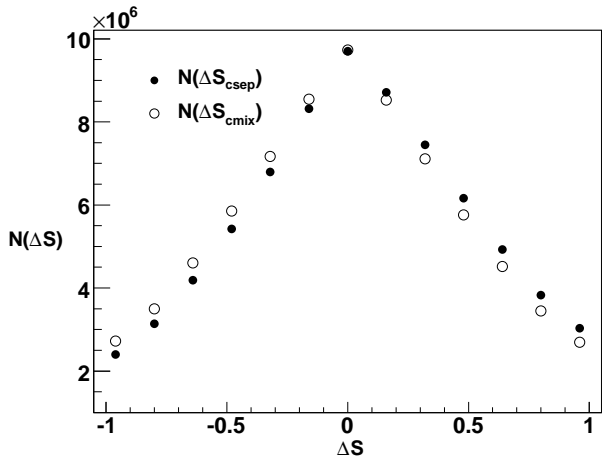}
\caption{$N(\Delta S)_{csep}$ and $N(\Delta S)_{cmix}$ distributions for 
simulations 
performed with $a1>0$ for all events.
}
\label{Fig2}
\end{figure}

\begin{figure}
\includegraphics[scale =0.45]{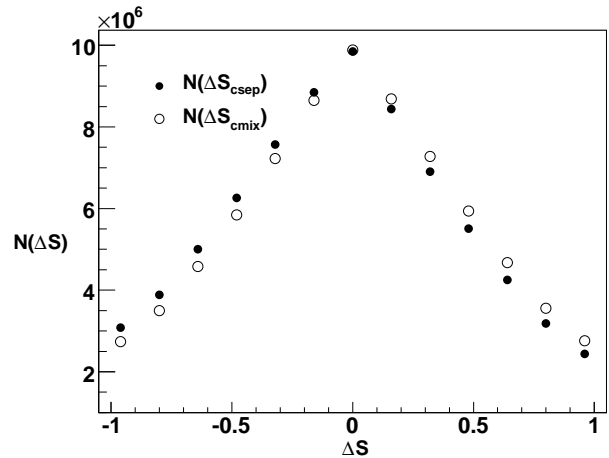}
\caption{$N(\Delta S)_{csep}$ and $N(\Delta S)_{mix}$ distributions for 
simulations 
performed with $a1<0$ for all events.
}
\label{Fig3}
\end{figure}

\begin{figure}
\includegraphics[scale =0.45]{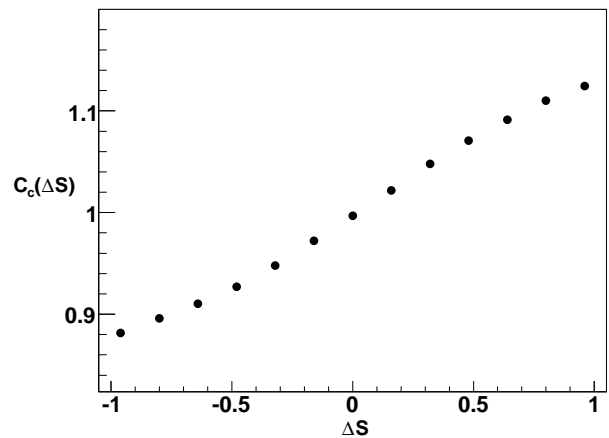}
\caption{Distribution for $C_c (\Delta S)$ obtained from the ratio of 
the distributions 
shown in Fig. \ref{Fig2}.
}
\label{Fig4}
\end{figure}

\begin{figure}
\includegraphics[scale =0.45]{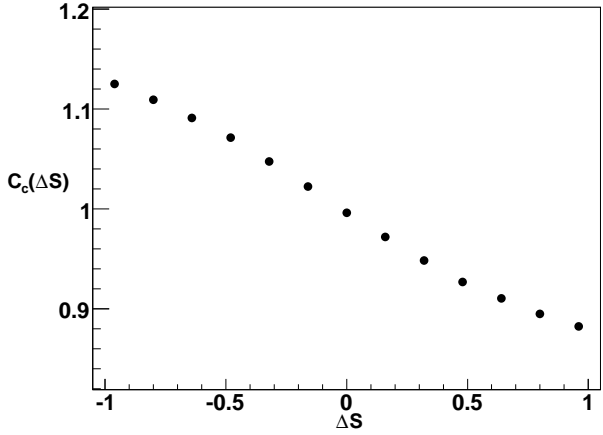}
\caption{$C_c (\Delta S)$ correlation function obtained from the ratio 
of the distributions 
shown in Fig. \ref{Fig3}.
}
\label{Fig5}
\end{figure}

\begin{figure}
\includegraphics[scale =0.45]{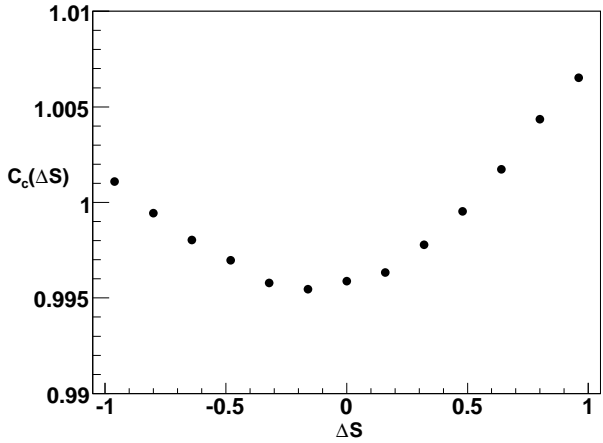}
\caption{$C_c (\Delta S)$ correlation function obtained with $a1>0$ in 
51\% of events 
and $a1<0$ in 49\% of events. 
}
\label{Fig8}
\end{figure}

\begin{figure}
\includegraphics[scale =0.45]{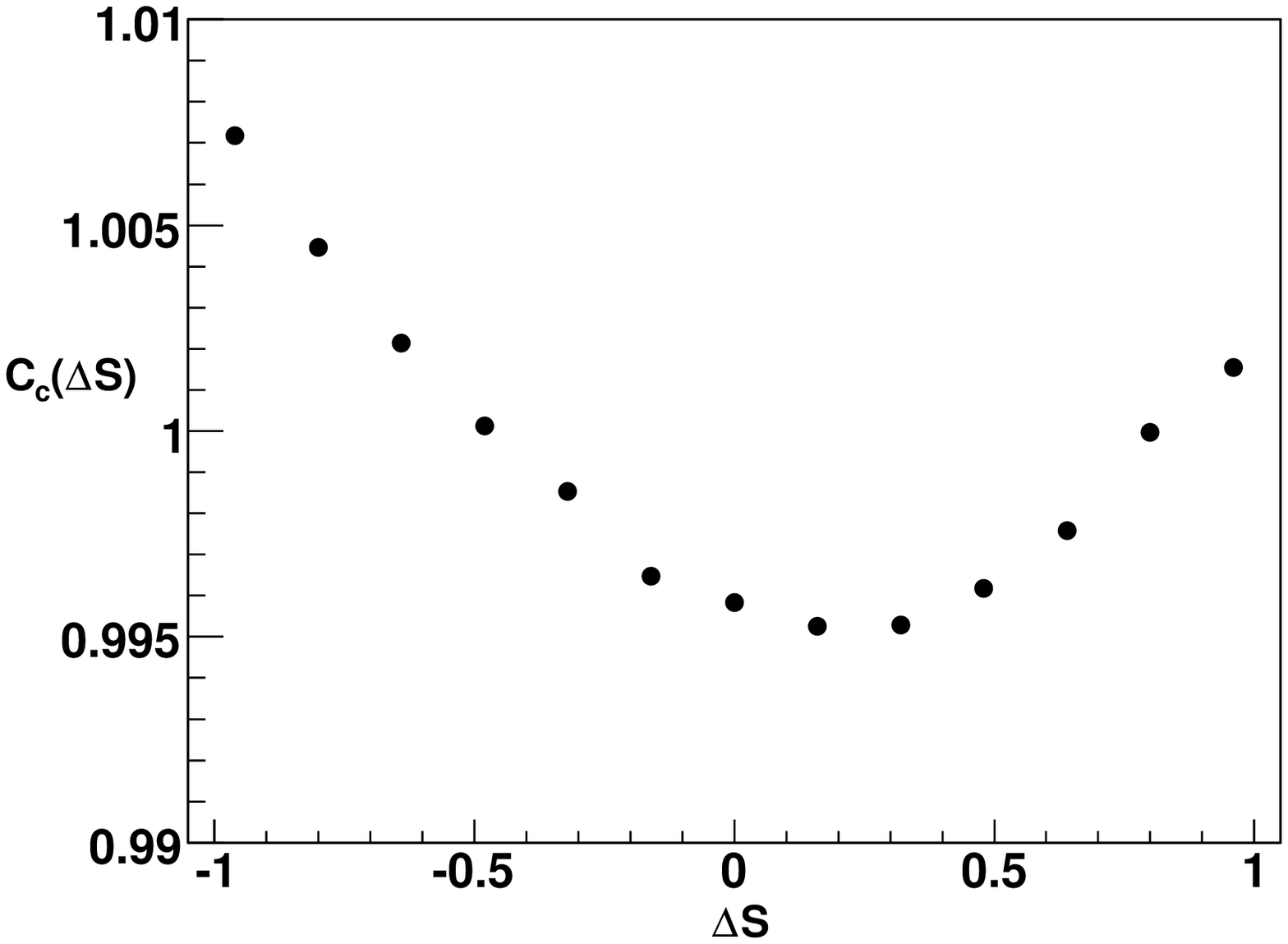}
\caption{$C_c (\Delta S)$ correlation function obtained with $a1>0$ in 
49\% of events 
and $a1<0$ in 51\% of events.
}
\label{Fig9}
\end{figure}

The distributions for $N(\Delta S)_{csep}$ 
(solid circles) and $N(\Delta S)_{mix}$ (open circles) are compared in Figs. \ref{Fig2} 
and \ref{Fig3} for $a_1 > 0$ and $a_1 <0$ (respectively) for all events. Fig. \ref{Fig2} 
shows that 
for $a_1 > 0$ the distribution for $N(\Delta S)_{csep}$ is shifted to 
the right when 
compared to 
that for $N(\Delta S)_{mix}$. Similarly Fig. \ref{Fig3} shows that for 
$a_1 <0$, the distribution for $N(\Delta S)_{csep}$ is shifted to the 
left 
when compared to that for $N(\Delta S)_{mix}$. 
%
%
%
Figs. \ref{Fig4} and \ref{Fig5} show the respective $C_c (\Delta S)$
distributions which result from the ratio of the distributions shown in 
Figs. \ref{Fig2} and \ref{Fig3}. They indicate sizable deviations from a flat distribution 
with positive and negative slopes respectively. Note that a flat distribution 
would be indicative of no charge-separation.
$C_c (\Delta S)$ distributions are shown in Figs. \ref{Fig8} and 
\ref{Fig9} for simulated events in which (i) 51\% of the events were 
generated with $a_1>0$ and the other 49 \% with $a_1<0$, 
and (ii) 49\% of the events with $a_1>0$ 
and 51\% of the events with $a_1<0$. In both cases, an asymmetric 
concave distribution is obtained but with opposite asymmetry.

%
%

\begin{figure}
\includegraphics[scale =0.45]{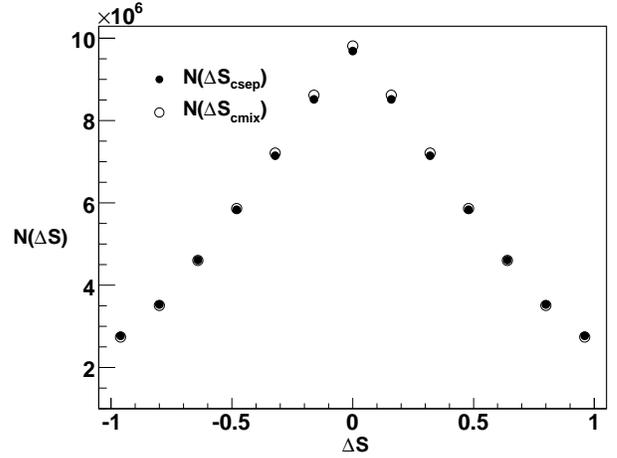}
\caption{Comparison of $N(\Delta S)_{csep}$ and $N(\Delta S)_{mix}$ 
distributions for simulations 
performed with $a1<0$ in 50\% of the events and $a1>0$ in the other 50\%.
}
\label{Fig6}
\end{figure}

\begin{figure}
\includegraphics[scale =0.45]{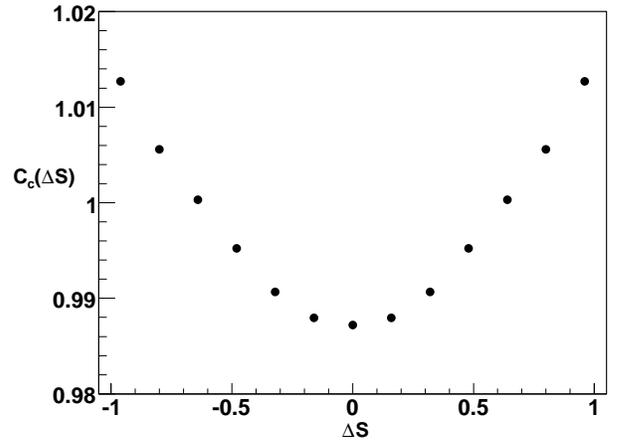}
\caption{$C_c (\Delta S)$ correlation function obtained from the ratio 
of the distributions 
shown in Fig. \ref{Fig6}.
}
\label{Fig7}
\end{figure}

	The distributions for $N(\Delta S)_{csep}$ and $N(\Delta 
S)_{mix}$ obtained for a simulation in which 50\% of the events were 
generated with $a_1 > 0$ and the other 50\% with $a_1<0$ are shown 
in Fig. \ref{Fig6}. This choice was made to mimic the effects of local 
parity violation implied by current models of topological charge generation 
in the QGP. For this scenario, Fig. \ref{Fig6} indicates that 
although the two distributions are strikingly similar, $N(\Delta S)_{csep}$ 
is slightly broader than $N(\Delta S)_{mix}$. 
This is made more transparent in Fig.\ref{Fig7} by the symmetric concave 
shape obtained for $C_c (\Delta S)$ from the ratio of these distributions.


\begin{figure}
\includegraphics[scale =0.45]{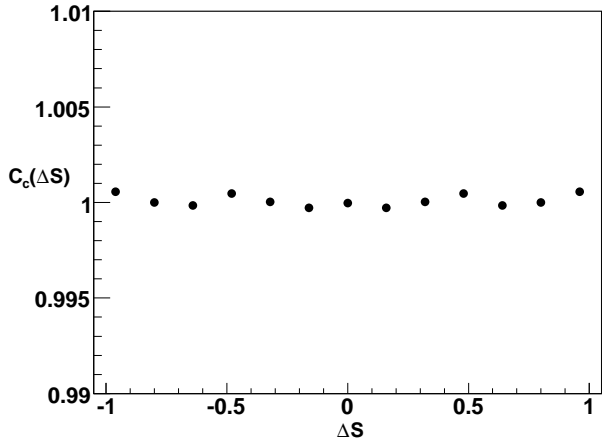}
\caption{$C_c (\Delta S)$ correlation function obtained for simulated 
events with 
only flow correlations.
}
\label{Fig10}
\end{figure}

\begin{figure}
\includegraphics[scale =0.45]{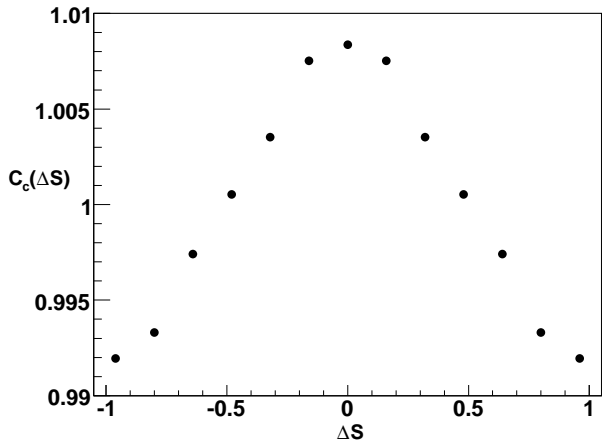}
\caption{$C_c (\Delta S)$ correlation function obtained for simulated 
events 
with flow on, resonance decay on and $a_1=0$.
}
\label{Fig11}
\end{figure}

	To investigate the influence of ``background'' correlations from flow,
jets and resonance decays, several simulations were performed with these 
correlations turned on or off. Fig. \ref{Fig10} shows the correlation function 
for a simulation in which only flow correlations are turned on for all events, 
i.e. $v_{2,4} \ne 0$, $a_1=0$ and resonance 
decays are turned off. The flat distribution indicated by Fig. \ref{Fig10}
shows that $C_c (\Delta S)$ is insensitive to flow.
In contrast to Fig. \ref{Fig10}, Fig. \ref{Fig11} shows a convex shape 
for $C_c (\Delta S)$ which results from the charge correlations 
associated with resonance decays which tends to bring opposite charges closer 
together in azimuth than on average. 
For this correlation function, the simulation was performed with  
flow on, $a_1=0$ and resonance decays on, for all events. Since these 
correlations have an opposite influence on the shape of $C_c(\Delta S)$ 
[compared to that for the charge separation signal], it is important to have 
the relative abundances of decay particles properly incorporated into 
the simulations. This is ensured by requiring the standard two particle 
opposite charge correlations from the simulations to match those for 
experimental data.
%

\begin{figure}
\includegraphics[scale =0.45]{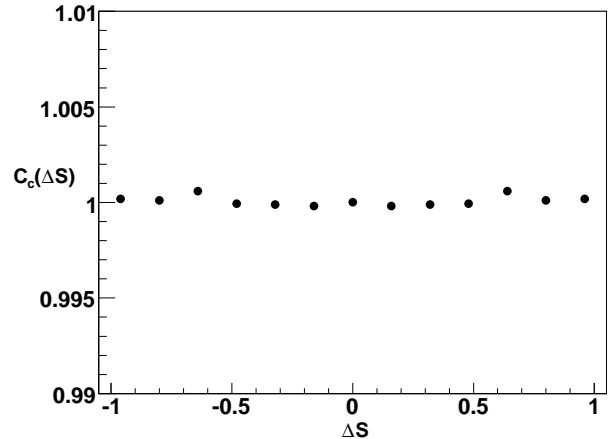}
\caption{$C_c (\Delta S)$ correlation function  obtained for simulated 
events 
with jets on, flow on, no resonance decay and $a_1$=0.
}
\label{Fig12}
\end{figure}

	The correlation function shown in Fig. \ref{Fig12} was obtained 
for simulated events in which flow is on, $a_1=0$, resonance decays are off, but 
jets are turned on for all events. It is very similar to the flat distribution seen in 
Fig. \ref{Fig10} and confirms the absence of any significant background correlations 
to $C_c (\Delta S)$ from jets. Because the same event is used to construct 
both $N(\Delta S)_{csep}$ and $N(\Delta S)_{mix}$ momentum correlation effects are 
also not expected to play any significant role. 

\begin{figure}
\includegraphics[scale =0.45]{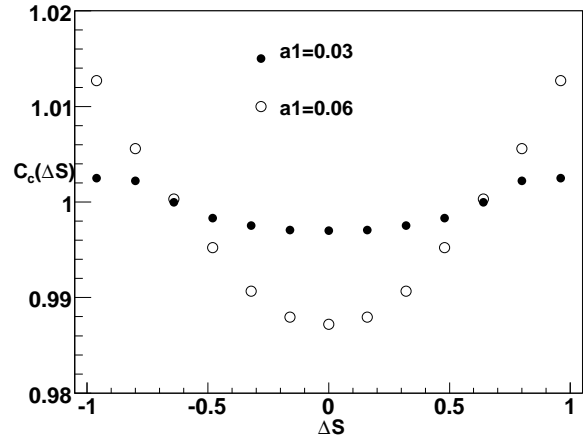}
\caption{$C_c (\Delta S)$ correlation functions  obtained for simulated 
events with $a_1$=0.03 and $a_1$=0.06.
}
\label{Fig13}
\end{figure}

	For an actual experimental correlation signal, the value of $a_1$ 
would be obtained by matching simulation to the observed correlation. 
The sensitivity of $C_c (\Delta S)$ to the the parameter used to specify 
the magnitude of the charge separation $a_1$ is demonstrated in Fig. \ref{Fig13}. 
It shows that $C_c (\Delta S)$ is responsive even to a relatively 
small charge asymmetry.

It is important to stress that the method presented here 
is very general. For this study it has been applied to the correlations 
between charges in an event. However, our methodology can be applied to the 
investigation of correlations within any observed particle property.

\section{Summary}
In summary, we have presented a new method which allows for good quantitative 
measurement of charge separation about the reaction plane. 
Our method involves the formulation  of a novel correlation  function 
$C_c (\Delta S)$ whose shape is concave only when  there is a non-zero charge 
separation signal. The strength of $C_c (\Delta S)$ is  related to the parameter $a_1$ 
which can be linked to a parity violating signal. It is noteworthy that 
experimentally a reaction plane obtained from second order flow ($v_2$) has a 
$\pm\pi$ ambiguity. Introduction of such an ambiguity in the 
simulations makes all correlations symmetric i.e. the sensitivity to 
global parity violation would be lost since the sign of $a_1$ cannot be 
determined. However the sensitivity of the correlation function to the  
magnitude of $a_1$ remains. 
Correlations due to flow, jets and momentum conservation, do not influence 
the shape nor the magnitude of $C_c (\Delta S)$. Therefore, $C_c (\Delta S)$ 
measurements could provide an important framework for investigating one of 
the most important problems of strong interaction 
theory -- the ``strong $\cal{CP}$ problem''.
%
%
%
\bibliography{Charge_sep_refs} 
\end{document}